\documentclass[12pt]{article}
\usepackage{graphicx}
\usepackage{float}
\usepackage{graphicx}
\begin{document}
\rightline{NKU-2017-SF2}
\bigskip


\newcommand{\be}{\begin{equation}}
\newcommand{\ee}{\end{equation}}
\newcommand{\noi}{\noindent}
\newcommand{\refb}[1]{(\ref{#1})}
\newcommand{\ra}{\rightarrow}
\newcommand{\bi}{\bibitem}
\newcommand{\bff}{\begin{figure}}
\newcommand{\eff}{\end{figure}}


\begin{center}
{\Large\bf Charged dilaton black hole in 2+1 dimensions as a particle accelerator}
\end{center}

\hspace{0.4cm}

\begin{center}
Sharmanthie Fernando \footnote{fernando@nku.edu}\\
{\small\it Department of Physics, Geology \& Engineering Technology}\\
{\small\it Northern Kentucky University\\
Highland Heights\\
Kentucky 41099\\
U.S.A.}\\

\end{center}

\begin{center}
{\bf Abstract}
\end{center}

In this paper we have studied particle collisions around a charged dilaton black hole in 2+1 dimensions. This black hole is a solution to the low energy string action in 2+1 dimensions. Time-like geodesics for charged particles are studied in detail. The center of mass energy for two charged particles colliding closer to the horizon is calculated and shown to be infinite if one of the particles has the critical charge.

\hspace{0.5cm} 



\section{Introduction}


Ban$\tilde{a}$dos, Silk and West \cite{ban} demonstrated that two particles colliding near the horizons of an extreme  Kerr black could create  large center of mass (CM) energy if one of the particles has critical angular momentum. This process is known as the BSW mechanism and has drawn many to study black holes in this context. Scattering energy of the particles in the vicinity of a rotating black hole can reach very large values for non extreme rotating black holes as demonstrated in  \cite{grib}.  The same is true for the Kerr-de Sitter black hole  \cite{li}.  BSW effect   of  the extreme Kerr-Taub-NUT space-time has been studied by Liu et.al \cite{liu}, of a rotating black hole in string theory (Sen black hole) were studied by Wei et.al. \cite{wei4},  and of a rotating cylindrical black hole was studied in  \cite{said}.

Non rotating charged black holes also could act as particle accelerators when they are extreme as pointed out by  Zaslavskill   \cite{oleg}. This phenomena where extreme charged black holes with charged particles acting as particle accelerators works for string black holes   was shown by Fernando  in \cite{fernando1} and for  Einstein-Maxwell-dilaton black holes  in \cite{mao}. Acceleration of particles also could occur inside spinning black holes as shown in  \cite{lake}.

Particle acceleration studies also have  been extended to regular black holes with a non-singular center: Rotating Hayward regular black hole as a particle accelerator was studied by Amir and Ghosh \cite{amir1}. Rotating Ay$\grave{o}$n-Beato-Garc$\grave{i}$a black hole which is also regular was studied in this context in \cite{amir2} and the rotating Bardeen black hole was  studied as a particle accelerator in \cite{amir3}. An excellent review on black  holes as particle accelerators is written by Harada \cite{harada}. Collisions of spinning particles also can create infinite center of mass energy as described in \cite{oleg2} \cite{guo} \cite{zhang}.

 In this paper our goal is to study particle acceleration in a charged dilaton black hole in 2+1 dimensions. Black holes in three dimensions originated from the well known BTZ black hole \cite{banados}\cite{ban2} which is a black hole with a negative cosmological constant. The BTZ black hole has attracted lot of attention since many aspects of black holes can be studied in a simpler setting. The dilaton black hole considered in this paper was derived by Chan and Mann \cite{chan1}. This particular black hole is important in the sense that it is a solution to low energy string theory in 2+1 dimensions. There are several extensions for dilaton black holes in 2+1 dimensions: New class of dilaton solutions were generated by applying T-dulity in \cite{chen}. Rotating dilaton solutions were generated by compactifying a 4D cylindrical solutions by Fernando \cite{fer1}. Dilaton black holes with non-linear electrodynamics were derived by Hendi et.al \cite{hendi}.
 
 As far as particle acceleration is considered, there are very few works in low dimensions. Particle acceleration by the BTZ black hole was studied by Yang et. al \cite{yang}, by the charged hairy black holes was studied by Sadeghi et.al. \cite{sad}, by the spinning dilaton black hole was studied by Fernando \cite{fer6}.

The paper is organized as follows: in section 2, we will present the details of the charged dilaton black hole in 2+1 dimensions. In section 3, time-like geodesics  for the charged dilaton black hole are presented.  The effective potential is analyzed   in section 4 and the center of mass is discussed in section 5. Finally the conclusion is given in section 6.


\section{ Introduction to the charged  dilaton  black hole in 2+1 dimensions}

The charged black hole considered in this paper was derived by Chan and Mann \cite{chan1}  for the following action,

\begin{equation} \label{action}
S = \int d^3x \sqrt{-g} \left[ R - \frac{B}{2} (\bigtriangledown \phi )^2 -
e^{-4 \alpha \phi} F_{\mu \nu} F^{\mu \nu} + 2 e^{\beta \phi} \Lambda \right]
\end{equation}
Here $\Lambda$ is the cosmological constant where $\Lambda > 0$ represents anti-de Siter and $\Lambda <0$ represents de-Sitter space.

The constants $\alpha$, $\beta$ and $B$ are arbitrary couplings, $\phi$ is the dilaton field, $R$ is the scalar curvature and $F_{\mu \nu}$ is the Maxwell's field strength. 
A family of black hole solutions with rotational symmetry for the above action were obtained as,

$$ds^2 = - \left(-2 M r ^{ \frac{2}{N} -1} + \frac{ 8 \Lambda r^2}{ (3N-2)N} + 
\frac{ 8 Q^2}{ (2 -N)N} \right)   dt^2$$ 
\begin{equation}
+  \frac{ 4 r^{\frac{4}{N}-2} dr^2}{N^2 \gamma^{4/N} \left((-2 M r ^{ 2/N -1} + \frac{ 8 \Lambda r^2}{ (3N-2)N}  + 
\frac{ 8 Q^2}{ (2 -N)N} \right) }
 + r^2  d \theta^2
\end{equation}
where,
\begin{equation}
k = \pm \sqrt{ \frac{N( 2 -N)}{ 2B}}; \hspace{1.0cm} 4\alpha k = \beta k = N-2; \hspace{1.0cm} 4\alpha = \beta;
\end{equation}
The dilaton field is given by
\begin{equation}
\phi = \frac{2 k}{N} ln \left( \frac{r}{r_0} \right)
\end{equation}

In this paper we will focus on a special class of black holes  with  $N=1, k= -\frac{1} {4}, \beta=4 \alpha=4$. In this case, the action in eq$\refb{action}$ is conformally coupled to the low energy string effective action in 2+1 dimensions. The black hole solution  in this particular case is given by,
\begin{equation} \label{metric}
ds^2= - f(r) dt^2 +  \frac{1}{h(r)} + r^2 d \theta^2
\end{equation}
where,
\be
f(r) =\left( -2Mr + 8 \Lambda r^2 + 8 Q^2 \right)
\ee
and
\be
h(r) =\frac{\left(-2Mr + 8 \Lambda r^2 + 8 Q^2 \right)  } {4 r^2}
\ee
The dilaton and the electric field are given by,
\begin{equation}
\phi = -\frac{1}{2}  ln \left(\frac{r}{r_0}\right) ; \hspace{1.0cm}F_{rt} = \frac{Q}{r^2}
\end{equation}
The above space-time has  two horizons if $M \geq 8 Q \sqrt{\Lambda}$ and they are  given by,
\begin{equation}
r_h =  \frac{M + \sqrt{ M^2 - 64 Q^2 \Lambda}}{8 \Lambda}; \hspace{1.0cm}
r_c = \frac{M - \sqrt{ M^2 - 64 Q^2 \Lambda}}{8 \Lambda}
\end{equation}
Here $r_h$ corresponds to the event horizon and $r_c$ corresponds to the Cauchy horizon. The two horizons merge to give an extreme black hole  when $M = 8 Q \sqrt{\Lambda}$. There  is a time-like singularity at $ r =0$. The  Hawking temperature $T_{Hawking}$ is given by,
\begin{equation}
T_{Hawking}= \frac{1}{4 \pi} |\frac{dg_{tt}}{dr}| \sqrt{-g^{tt} g^{rr}} |_{r=r_h} = \frac{M}{4 \pi r_h} \sqrt{ 1 - \frac{64 Q^2 \Lambda}{M^2}}
\end{equation}

There are several works on the charged dilaton black hole presented above: Quasinormal modes of charged and neutral scalar fields were studied by Fernando  \cite{fer2}, \cite{fer3} and by Myung et.al. \cite{my}. Grey body factors of neutral scalar fields were studied in \cite{fer4}. Geodesics of the black hole were studied by Fernando et.al. in  \cite{fer5}.


\section{ Time-like geodesics of the test particles of the charged dilation black hole}


In this section we will present  time-like geodesics of a test particle with charge per unit mass $e$ around the black hole considered in this paper. We will follow the  formalism in Chandrasekhar's book  \cite{chandra}.  The Lagrangian of the charged test particle in this black hole back ground is given by,
\be \label{lag}
{\cal{L}} =  -\frac{1}{2} \left( g_{\mu \nu} \frac{ dx^{\mu}}{d \tau} \frac{ dx^{\nu}}{ d \tau} \right)  + e A_{\mu} \frac{ dx^{\mu}}{d \tau}
\ee
Here the parameter $\tau$ is the proper time  of the charged particle. The metric functions and $A_{\mu}$ are given by,
\be
g_{tt} = -f(r); \hspace{1 cm} g_{rr} = \frac{1}{ h(r)}; \hspace{1 cm} g_{\phi \phi} = r^2; \hspace{1 cm} A_{t } =  \frac{Q}{r}
\ee
Now, the Lagrangian can be written explicitly as,
\be \label{lag2}
L = - \frac{1}{2} \left( - f(r) \left( \frac{dt}{d\tau} \right)^2 +  \frac{1}{h(r)}\left( \frac{dr}{d \tau} \right)^2 + r^2 \left(\frac{d \phi}{d \tau} \right)^2  - 2 \frac{e Q}{r}  \left(\frac{ dt}{ d \tau} \right) \right)
\ee
Each coordinate in the space-time has a corresponding canonical momenta $p_{\mu}$  which can be obtained from the Lagrangian as,
\begin{equation} \label{pt}
p_t =  \frac{\partial L}{\partial\dot{ t}} = f(r) \dot{t}  + \frac{e Q}{r}
\end{equation}
\begin{equation} \label{phi}
p_{\phi}  =    -   \frac{\partial L}{\partial\dot{ \phi}}=    r^2 \dot{\phi} 
\end{equation}
\begin{equation}
p_{r} = - \frac{\partial L}{\partial \dot{r}}  =  \frac{\dot{r}} { h(r)}
\end{equation}

The charged  dilaton  black hole  has two Killing vectors $\partial_t$ and $ \partial_{\phi}$. Hence, the canonical momenta  $p_t$ and $p_{\phi}$ are conserved and  these constants    are labeled as energy per unit mass $E$ and angular moment per unit mass $L$.   From eq.$\refb{pt}$ and eq.$\refb{phi}$,  $E$ and $L$  can be solved to be,
 \begin{equation} \label{pt2}
E = f(r) \dot{t} +  \frac{ e Q}{r} 
\end{equation}
\begin{equation} \label{pphi2}
L=   r^2  \dot{\phi} 
\end{equation}
Eq$\refb{pt2}$ and eq$\refb{pphi2}$ can be  solved to obtain $\dot{t}$ and $\dot{\phi}$ as,
\be \label{tdot4}
\dot{t} = \frac{  \left( E - \frac{e Q}{r} \right)} { f(r)} = u^t
\ee
\be \label{pdot4}
\dot{\phi} = \frac{  L}{r^2} = u^{\phi}
\ee
We will assume that $\dot{t} > 0$ for all $r > r_h$. Hence   the motion is forward in time outside the black hole event horizon. This implies,
\be \label{con}
 \left(E  - \frac{e Q}{r} \right)  > 0,  \hspace{1 cm} \forall \hspace{0.1 cm}  r > r_h
\ee

The components of the velocity of the particle are given by  $u^{\mu} = \frac{ dx^{\mu}}{d \tau}$ and they are normalized as, $u^{\mu} u_{\mu} = -1$. The normalized condition is expressed as
\be \label{nor}
g_{tt} (u^t)^2 + g_{rr}(u^r)^2 + g_{\phi \phi} (u^{\phi})^2  = -1
\ee
leading to,
\be \label{rdot}
 \dot{r}^2   =  \frac{ - \left( 1 + g_{tt} \dot{t}^2 + g_{\phi \phi} \dot{\phi}^2  \right) }{ g_{rr} } 
\ee
By substituting $u^t$ and $u^{\phi}$ from eq$\refb{tdot4}$ and eq$\refb{pdot4}$, one can obtain   $u^r= \dot{r}$ as,

\be \label{rdot}
\dot{r}^2 = \frac{1}{ 4  r^4} \left( -  f(r) ( r^2 + L^2) + ( E r - eQ)^2\right)
\ee
By identifying,
\be
K = ( E r - eQ)
\ee
and
\be
H^2 = K^2 - f(r) ( r^2 + L^2)
\ee
$\dot{r}$ can be written in a short form as,
\be
\dot{r} = -  \frac{ H}{ 2 r^2} = u^r
\ee
The minus sign for $\dot{r}$  is chosen since we assume the particle is  falling towards the black hole.


\section{ Effective potential for the charged dilaton black hole}

\begin{figure} [H]
\begin{center}
\includegraphics{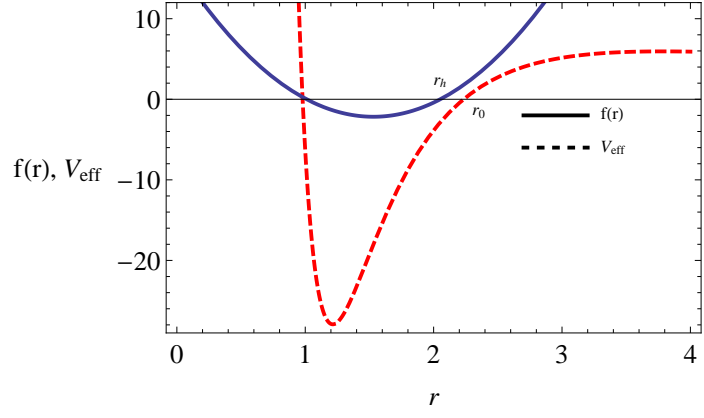}
\caption{The figure shows  $f(r)$ and $V_{eff}$ vs $r$. Here $ M = 12.25, Q = 1.44, E = 7.7, \Lambda =1, e = 1.1$ and $ L = 11.5$}
\label{pot1}
\end{center}
\end{figure}

\begin{figure} [H]
\begin{center}
\includegraphics{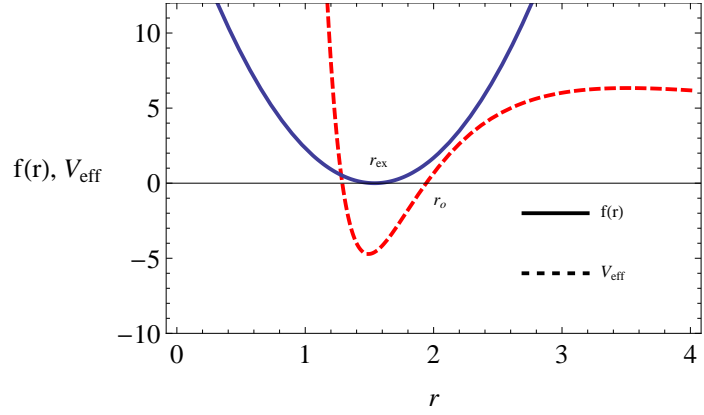}
\caption{The figure shows  $f(r)$ and $V_{eff}$ vs $r$ for an extreme black hole. Here $ M = 12.32, Q = 1.54, E = 7.7, \Lambda =1, e = 1.1$ and $ L = 11.5 $}
\label{pot2}
\end{center}
\end{figure}

\begin{figure} [H]
\begin{center}
\includegraphics{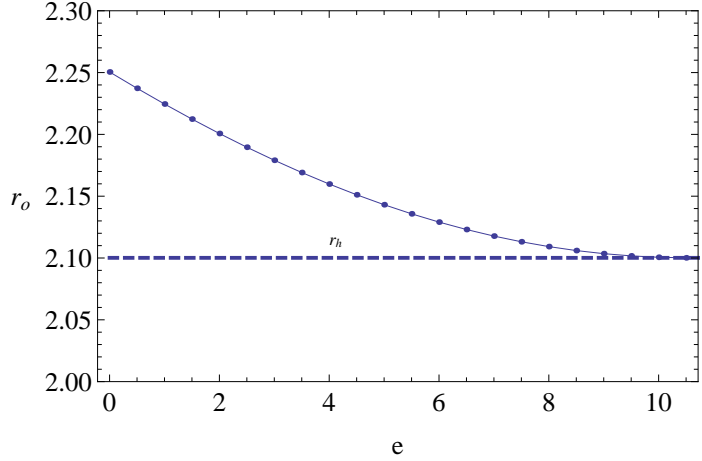}
\caption{The figure shows  $r_0$ vs $e$. Here $ M = 12.35, Q = 1.44, E = 7.6, \Lambda =1$ and $ L = 13.1 $}
\label{roots}
\end{center}
\end{figure}

In this paper our main  motivation is to understand the collision near the horizon. Hence it is vital to see if the particle could reach the horizon. In order to achieve that we will study the effective potential for the charged particle given by the relation,
\be 
\dot{r}^2  + V_{eff} =0
\ee
From  eq$\refb{rdot}$, one obtain $V_{eff}$ as,
\be
V_{eff} =  \frac{  f(r) ( r^2 + L^2) - K^2} { 4 r^4}
\ee
For $r \ra \infty $, $V_{eff} \ra 2 \Lambda$ and for $ r \ra 0$, $V_{eff} \ra \infty$. Since $\dot{r} = \sqrt{- V_{eff} }$, the particle motion is possible only  in the regions where $V_{eff} < 0$. Hence the particle should start the motion from a finite distance from the black hole. In Fig$\refb{pot1}$ and Fig$\refb{pot2}$, $V_{eff}$ and $f(r)$ are plotted. Fig.$\refb{pot1}$ represents a non-degenerate black hole and Fig$\refb{pot2}$ represents an extreme black hole. The effective potential has two roots for the given parameters. What is noteworthy is that the larger root lies outside the event horizon.  At $ r = r_h$,
\be
V_{eff} = -  \frac{ ( E r_h - e Q)^2}{ 4 r_h^4} = - \frac{ K(r_h)^2}{ 4 r^4}
\ee
Hence $V_{eff} \leq 0 $ at the horizon. $V_{eff} =0$ for a critical charge $e_c$ given by,

\be
e_c = \frac{ E r_h}{ Q} = \frac {E}{ Q} \left( \frac{ M + \sqrt{ M^2 - 64 Q^2 \Lambda}}{ 8 \Lambda} \right)
\ee
Notice that when $V_{eff}=0$, $K(r_h) =0$. We calculated the roots of $V_{eff}$ numerically and have plotted in  Fig$\refb{roots}$. It is clear that the root  $r_o \geq r_h$ for all $ e > e_c$. 
Therefore  if $ e > e_c$, the particle can start at rest from a finite distance  from the horizon and fall towards the black hole. The larger the $e$, further away the particle can start its motion.

\section{ Collision of two charged particles and their center of mass energy}

\begin{figure} [H]
\begin{center}
\includegraphics{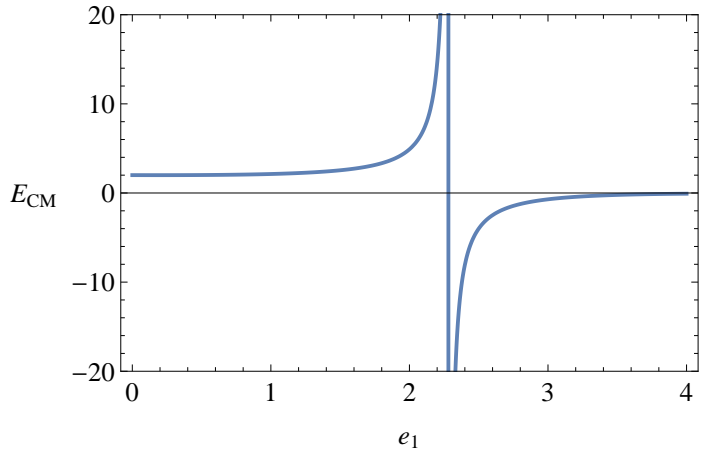}
\caption{The figure shows  $E_{cm}$ vs $e_1$. Here $ M = 10, Q = 1, \Lambda =1, E_1 = 1.14, E_2 = 0.725, L_1 = 0.4, L_2 = 0.3$}
\label{ecm}
\end{center}
\end{figure}

In this section we will calculate the center of mass energy of two charged particles colliding closer to the horizon. The four velocities of the two particles are,  $u_1^{\mu}$ and $u_2^{\mu}$. The rest mass of the particles are assumed to be equal and is given by $m_0$. Hence the center of mass energy is given by,
\be
\tilde{E}_{cm} = 2 m_0^2 ( 1 - g_{\mu \nu} u_1^{\mu} u_2^{\nu} )
\ee
In the rest of the paper we will calculate $E_{cm} = \tilde{E}_{cm}/(2 m_0^2)$ given by,
\be
E_{cm} = \frac{ \left( f(r) ( r^2 - L_1 L_2) - ( K_1 K_2 - H_1 H_2) \right)} { r^2 f(r)}
\ee
where,
\be
K_1 =  r E_1 - e_1 Q
\ee
\be
K_2 = r E_2 - e_2 Q
\ee
\be
H_1 = K_1^2 - f(r) ( L_1^2 + r^2)
\ee
\be
H_2 = K_2^2 - f(r) ( L_2^2 + r^2)
\ee
When the particles reach  the horizon with  $ r = r_{h}$, $f(r) \ra 0$, $ H_1 \ra \sqrt{K_1^2}$, and $ H_2 \ra \sqrt{K_2^2}$. Therefore,
\be \label{energy}
E_{cm}(r \ra  r_h ) = \frac{ 1}{ r_h^2 f(r_h) } \left( K_1 K_2 - \sqrt{K_1^2}  \sqrt{K_2^2} \right)
\ee
Since $f(r_h)=0$, the denominator of the above expression is zero. Also, due to the condition given in eq$\refb{con}$, $ K_1(r_h), K_2(r_h) \geq 0$. Hence the numerator is also zero leading to an undetermined value for the center of mass energy at the horizon. However, one can use the L' Hospital's  rule to calculate the limiting value of  $E_{cm} ( r \ra r_h)$ resulting,

\be \label{limit}
E_{cm} = \frac{  A_1 r_h^4 + A_2 r_h^3 + A_3 r_h^2 + A_4 r_h + A_5}{ 2 r_h^2 K_1(r_h) K_2(r_h)}
\ee
where
\be
A_1 = -(E_1 - E_2)^2 
\ee
\be
A_2 = 2 ( e_1 - e_2) ( E_1 - E-2) Q
\ee
\be
A_3 = - ( E_1 L_2 + E_2 L_1)^2 - (e_1 - e_2)^2 Q^2
\ee
\be
A_4 = 2 Q( e_2 L_1 + e_1 L_2) ( E_2 L_1 + E_1 L_2) 
\ee
\be
A_5 = - Q^2( e_2 L_1 + e_1 L_2)^2 
\ee
The numerator of eq$\refb{limit}$ is finite. So, if $K_1(r_h) =0$ or $K_2(r_h) =0$, the center of mass energy becomes infinity. When one solve $K_1(r_h) =0$ or $K_2(r_h) =0$,  one obtain the critical charge
$e_{c1}$ or $e_{c2}$ which was discussed in section(4).  We have plotted $E_{cm}$ in Fig$\refb{ecm}$  for varying $e_1$ with all other parameters fixed. 
When particle number 1 reach the critical charge $e_1c$, $E_{cm}$ goes to infinity. Notice that the critical charge  has to be reached from the left of the asymptotic value in order to get a positive value for the center of mass energy.

It is important to notice that to achieve high values of $E_{cm}$, only one of the particles should have the critical charge. If both have the critical charges, then $E_{cm}$ will be finite. One can prove it as follows: When both particles have critical charge, both $K_1(r_h) = K_2(r_h) =0$. This implies that  $H_1(r_h) = H_2(r_h) =0$. Hence for this special case,
\be
E_{CM} = \frac{ \left(  r_h^2 - L_1 L_2  \right)} { r_h^2} = 1 - \frac{ L_1 L_2} { r_h^2}
\ee
which is  finite.

\section{Conclusions}

In this paper we have studied particle collisions around a charged dilaton black hole in 2+1 dimensions. This black hole is a solution to the low energy string theory in 2+1 dimensions. The given black hole has two horizons which depend on $M, Q$ and $\Lambda$. If $ M = 8 Q \sqrt{\Lambda}$, it has degenerate horizon.

The time-like geodesics for  charged particles are studies in detail. The three velocities, $u^t, u^r, u^{\phi}$ are calculated. The effective potential  reaches $2 \Lambda$ for large $r$ and goes to infinity for $ r \ra 0$. When the parameters are such that there are horizons, the effective potential has a root $r_0$ larger than the event horizon $r_h$. Hence a particle could start the  motion at rest with $r < r_0$ and fall into the black hole. There is a critical charge at which $ r_0 = r_h$.

The main goal in this paper was to study the particle collision around the black hole. We found that the center of mass energy of two charged particles colliding will becomes infinity if one of the charged particles has the critical charge. Hence the charged dilaton black hole  can be a particle accelerator.



\end{document}